\documentclass[lettersize,journal]{IEEEtran}

\usepackage{amsmath,amssymb,amsfonts}
\usepackage{algorithm}
\usepackage{algorithmic}
\usepackage{array}
\usepackage{tabularx}
\usepackage[caption=false,font=normalsize,labelfont=sf,textfont=sf]{subfig}
\usepackage{textcomp}
\usepackage{stfloats}
\usepackage{url}
\usepackage{verbatim}
\usepackage{graphicx}
\usepackage{cite}
\usepackage{scalerel}
\usepackage{tikz}
\usetikzlibrary{svg.path}

\usepackage{booktabs}

\definecolor{orcidlogocol}{HTML}{A6CE39}
\tikzset{
orcidlogo/.pic={
\fill[orcidlogocol] svg{M256,128c0,70.7-57.3,128-128,128C57.3,256,0,198.7,0,128C0,57.3,57.3,0,128,0C198.7,0,256,57.3,256,128z};
\fill[white] svg{M86.3,186.2H70.9V79.1h15.4v48.4V186.2z}
svg{M108.9,79.1h41.6c39.6,0,57,28.3,57,53.6c0,27.5-21.5,53.6-56.8,53.6h-41.8V79.1z M124.3,172.4h24.5c34.9,0,42.9-26.5,42.9-39.7c0-21.5-13.7-39.7-43.7-39.7h-23.7V172.4z}
svg{M88.7,56.8c0,5.5-4.5,10.1-10.1,10.1c-5.6,0-10.1-4.6-10.1-10.1c0-5.6,4.5-10.1,10.1-10.1C84.2,46.7,88.7,51.3,88.7,56.8z};
}}

\newcommand\orcidicon[1]{\href{https://orcid.org/#1}{\mbox{\scalerel*{
\begin{tikzpicture}[yscale=-1,transform shape]
\pic{orcidlogo};
\end{tikzpicture}
}{||}}}}
\usepackage{hyperref}

\begin{document}

\title{An Embedded Industrial Irrigation System\\ with Weather-Integrated Intelligent Scheduling}

\author{Kiarash Zangooei, Arya Masoodpay, and Hamid Reza Ghaeini\orcidicon{0000-0003-2614-7900}, \IEEEmembership{Professional Member, IEEE}
\thanks{Kiarash Zangooei, Arya Masoodpay, and Hamid Reza Ghaeini are with the College of Engineering, University of Tehran, Tehran, Iran.}
\thanks{Corresponding author: Hamid Reza Ghaeini (ghaeini@ut.ac.ir)}}

\markboth{IEEE}%
{Zangooei \MakeLowercase{\textit{et al.}}: An Embedded Industrial Irrigation System with weather-aware irrigation scheduling}

\maketitle

\begin{abstract}
Efficient irrigation management requires timely responses to soil moisture conditions and anticipated weather changes. This article presents an embedded, weather-based irrigation system that integrates soil moisture monitoring using multiple sensors, edge-based decision-making, automated pump control, and local data management. Four capacitive soil moisture sensors are connected to an STM32F103C8T6 microcontroller for data acquisition, while a BTT Pi V1.2 platform evaluates the sensor data, retrieves short-term weather forecasts, makes irrigation decisions, manages a graphical user interface, and stores operational data locally. Individual sensor readings serve as the basis for zone-level irrigation decisions, while the median of four time-averaged readings supports monitoring and analysis. Forecasted precipitation is taken into account as effective precipitation to reduce the calculated irrigation requirement before the pump is activated. The system was evaluated as part of a one-month cultivation trial with radishes, during which a total of 38,112 data points from measurement cycles were recorded. The median value of normalized soil moisture averaged 47.93\% and remained predominantly within the configured control range. Three irrigation events with a planned total irrigation volume of 26.68 L were recorded. Finally, we demonstrate the technical feasibility of integrating sensors, weather data, local data processing, automated control, and data collection into a scalable platform for precision irrigation.
\end{abstract}

\begin{IEEEkeywords}
Edge Computing, Precision Agriculture, Smart Irrigation, Soil Moisture Measurement, Weather Forecasting.
\end{IEEEkeywords}

\section{Introduction}
\label{sec:introduction}

\IEEEPARstart{W}{ater} management in agriculture is increasingly critical due to rising water demand, climatic fluctuations, and the need for sustainable food production \cite{unwater2020climate,postel1996human}. Fixed-schedule irrigation often fails to account for dynamic conditions such as soil moisture fluctuations, precipitation events, and changing atmospheric conditions \cite{macadam2020irrigation}, leading to over- or under-irrigation \cite{rai2017efficiency}. Smart irrigation strategies that adapt to real-time, on-site conditions are therefore essential for water-use efficiency and sustainable agriculture \cite{mahajan2025precision}.

Precision agriculture and IoT technologies have enabled automated irrigation based on environmental sensors, wireless communication, and embedded computer systems. Soil moisture sensors are widely used because they provide direct information about the water status of the growing medium \cite{muzdrikah2018calibration} and make it possible to overcome the limitations of time-based irrigation schedules through control loops \cite{khasawneh2026optimizing}. Various strategies have been investigated, including control based on moisture thresholds, evapotranspiration-based scheduling, and weather-based management \cite{navarrohellin2016decision}. Moisture-based methods directly reflect plant growth conditions but do not account for impending precipitation, while weather-based approaches can reduce unnecessary irrigation but depend on the accuracy of forecasts and a reliable computational infrastructure \cite{sui2017irrigation}. Combining both approaches enables more adaptive irrigation decisions.

Despite this progress, many existing solutions are based on cloud architectures that limit their use in remote areas, focus on algorithms without full embedded implementations, or lack practical, long-term validation in the field. To address these gaps, this article presents an embedded, weather-based irrigation system that combines four capacitive soil moisture sensors and an STM32F103C8T6 microcontroller for data acquisition with a BTT Pi V1.2 edge platform, which handles median-based sensor fusion, irrigation decision-making, weather data retrieval, graphical user interface operation, and database management.

The main contributions of this work are:

\begin{itemize}
\item A fully integrated irrigation architecture that has been experimentally evaluated and combines multisensor soil moisture measurement, edge computing, weather forecast data, automated control, and data management.
\item A median-based sensor fusion method on the BTT Pi V1.2 for determining a representative soil moisture value from four independent capacitive sensors.
\item A weather-based decision strategy that uses short-term forecasts to postpone unnecessary irrigation before predicted precipitation.
\item A practical experimental evaluation using radish plants, which provides a realistic dataset on soil moisture, irrigation events, and weather information.
\end{itemize}

The remainder of this article is structured as follows: Section~\ref{sec:architecture} introduces the system architecture. Section~\ref{sec:methodology} describes the methodology for irrigation control. Section~\ref{sec:implementation} deals with the hardware and software implementation. Section~\ref{sec:results} presents the experimental setup and the results. Section~\ref{sec:analysis} contains a detailed analysis. Section~\ref{sec:related_work} provides an overview of related work. Section~\ref{sec:discussion} discusses the results, and Section~\ref{sec:conclusion} concludes the paper.

\section{System Architecture}
\label{sec:architecture}

\begin{figure}
\centering
\includegraphics[width=\linewidth]{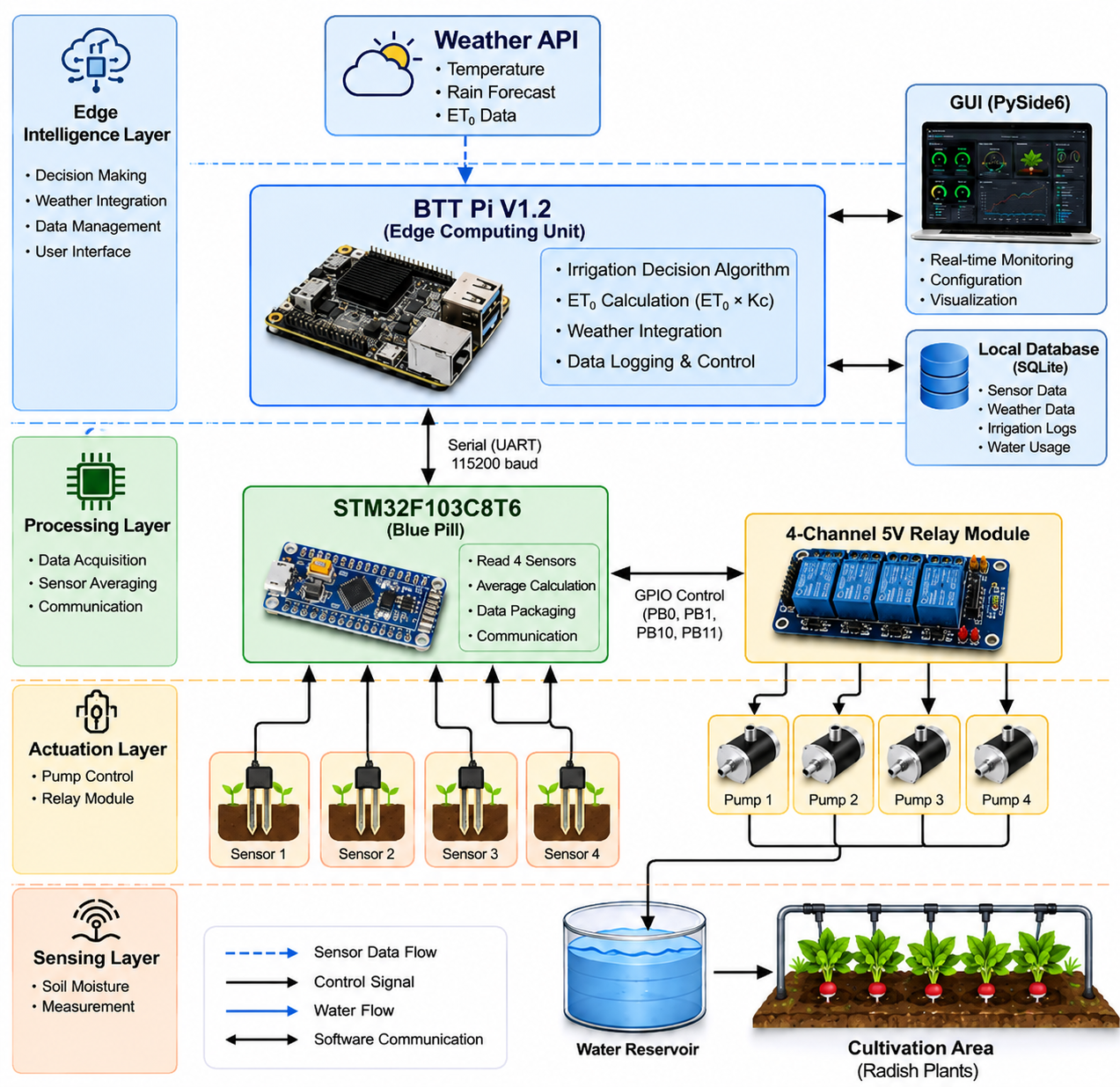}
\caption{Overall architecture of the proposed weather-dependent embedded irrigation system.}
\label{fig:architecture}
\end{figure}

The proposed system is an embedded cyber-physical architecture that integrates environmental sensing, edge-based decision-making, automated control, and user interaction. Unlike conventional systems, which rely on fixed schedules or isolated measurements, it combines real-time soil moisture data with short-term weather forecasts. The architecture consists of three layers—sensing, processing/decision-making, and control/monitoring—as shown in Figure~\ref{fig:architecture}.

The data acquisition layer (four capacitive sensors and the STM32F103C8T6) forwards preprocessed measurement values to the processing layer (BTT Pi V1.2), which combines the sensor data with forecasts from the weather API to generate irrigation commands. These commands control the relay-controlled pumps and update both the database and the graphical user interface in the control/monitoring layer.

\subsection{Sensor Layer}

Four capacitive soil moisture sensors, distributed across the cropland to detect spatial variations, provide data to the STM32F103C8T6 for local preprocessing. Since individual sensors may vary due to soil heterogeneity, placement, and noise, a median-based fusion approach provides a representative value for monitoring \cite{anbarasu2026multiparameter,wang2016datafusion}, while individual measurements drive zone-level control.

\subsection{Edge Computing and Decision-Making Level}

The BTT Pi V1.2 executes the irrigation algorithm, retrieves weather forecasts, manages data, and provides the graphical user interface (GUI). It receives the four normalized moisture values and combines them with short-term forecasts from an online weather API. If the moisture content in a zone falls below the threshold, the controller calculates the effective precipitation based on the forecast and activates irrigation only if the net demand remains positive. This decentralized design reduces the computational load on the sensor nodes and enables flexible control updates at the edge, thereby improving responsiveness \cite{shi2016edge,shakeri2018uav}. Weather data is cached in memory once per interval (default: one hour); without a persistent cache or the use of simulated data, API errors can interrupt the decision-making cycle.

\subsection{Execution and Monitoring Levels}

The edge controller sends pump commands via a serial interface to the STM32F103C8T6, which controls the relay module and the individual pumps. In addition to automated irrigation, the system stores environmental measurements, weather data, irrigation operations, and status parameters in a local database and features a graphical user interface for visualization and configuration. Irrigation decisions are based on sensor readings taken prior to irrigation, and pump operation is time-controlled without real-time feedback during irrigation.

\section{Methodology}
\label{sec:methodology}

\begin{figure}
\centering
\includegraphics[width=0.5\textwidth]{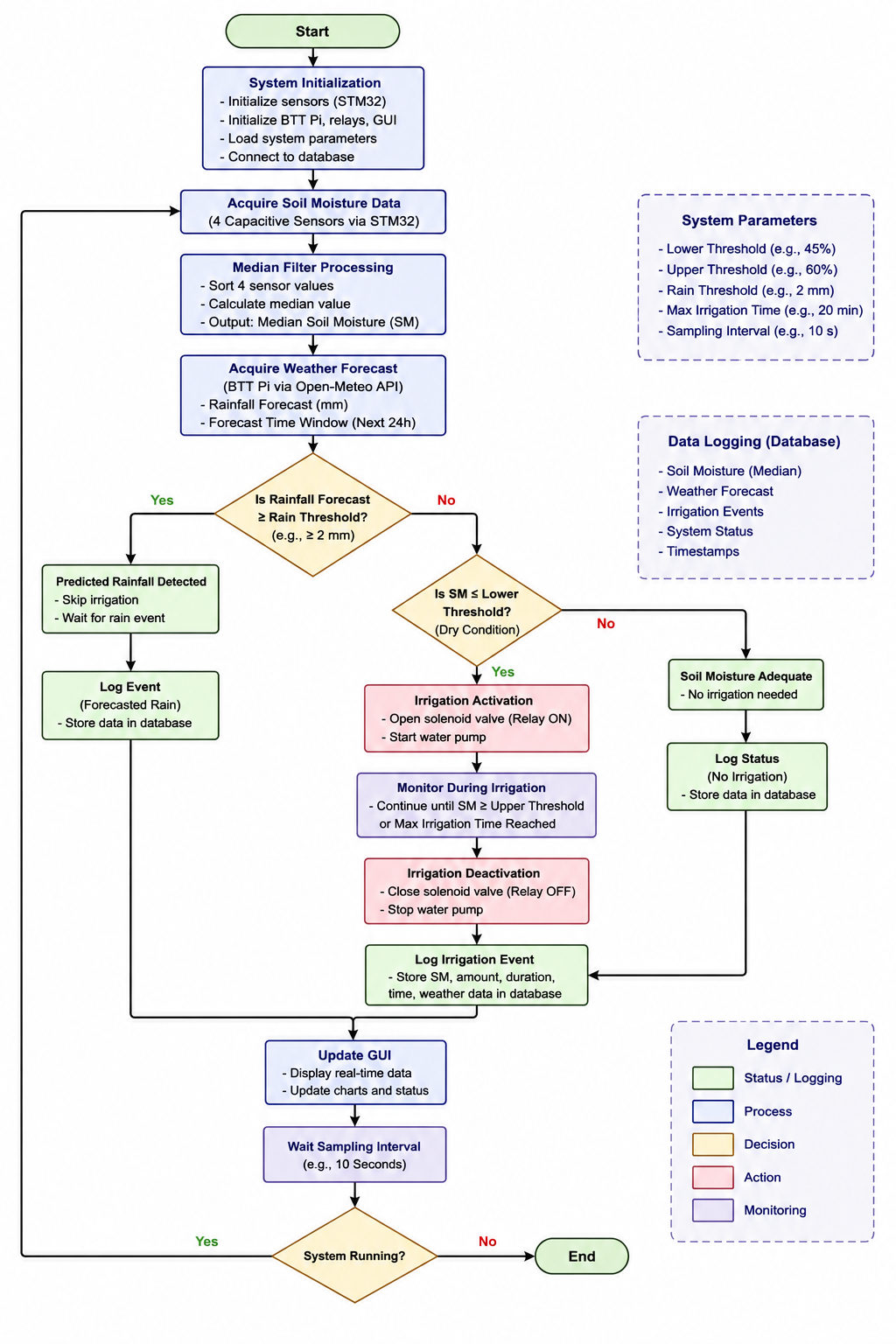}
\caption{Flowchart of the proposed weather-dependent decision algorithm for irrigation.}
\label{fig:algorithm_flowchart}
\end{figure}

The proposed methodology combines moisture monitoring using multiple sensors, local sensor fusion, weather forecasts, and automated pump control to plan irrigation based on actual soil conditions while avoiding unnecessary irrigation when rain is expected. The STM32F103C8T6 acquires the four measured values and sends them to the BTT Pi V1.2, which handles sensor fusion, retrieving weather data, decision-making, database management, and operation of the graphical user interface, as shown in the flowchart in Figure~\ref{fig:algorithm_flowchart}.

In each cycle, the average of the four measured values is calculated, the median is determined, and a check is performed for each zone to see if the soil moisture is below the threshold value; if this is the case, the forecast is retrieved, the effective precipitation and the net requirement are calculated, and irrigation and pump runtime are activated only if the net requirement is positive.

\subsection{Measuring Soil Moisture with Multiple Sensors}

Four capacitive sensors, chosen for low cost, durability, and embedded compatibility \cite{muzdrikah2018calibration}, are sampled over a 60-second window on the BTT Pi V1.2 to produce time-averaged values $\bar{M}_i$, $i\in\{1,2,3,4\}$. After each cycle, the system waits a configurable interval (default 300 s), giving a total cycle length of about 360 s. With values sorted as $\bar{M}_{(1)}\leq\bar{M}_{(2)}\leq\bar{M}_{(3)}\leq\bar{M}_{(4)}$, the median is

\begin{equation}
\bar{M}_{\mathrm{median}} = \frac{\bar{M}_{(2)}+\bar{M}_{(3)}}{2}.
\label{eq:soil_median}
\end{equation}

This value is used for monitoring and analysis, while individual average values control the system at the zone level.

\subsection{Information on Crops and Irrigation Requirements}

The system supports configurable crop-specific parameters; radishes were used in this study. The cultivation area is divided into four zones of equal size ($A_{\mathrm{zone}} = A_{\mathrm{total}}/4$), each equipped with a sensor and a pump. Given an effective soil depth $d$, lower and upper moisture thresholds $L, U$, a target value $T=(L+U)/2$, and a soil water capacity coefficient $C_w$, the crop’s evapotranspiration is

\begin{equation}
ET_c = K_c \cdot ET_0,
\label{eq:crop_et}
\end{equation}

and the zone irrigation requirement is \cite{pereira2021cropcoefficients}

\begin{equation}
D_{\mathrm{req}} = A_{\mathrm{zone}} \cdot d \cdot 1000 \cdot C_w \cdot (T - M_i)/100 + ET_c \cdot A_{\mathrm{zone}},
\label{eq:irrigation_requirement}
\end{equation}

where $M_i$ is the current zone moisture. The irrigation efficiency $\eta_{\mathrm{irr}}$ is applied only after accounting for effective precipitation.

\subsection{Weather-Based Irrigation Planning}

The BTT Pi V1.2 retrieves a two-day forecast, using precipitation and $ET_0$ for irrigation calculations; temperature is recorded/displayed but not used directly. Forecasted precipitation $R_f$ (mm), efficiency factor $\eta_r$, and zone area give effective precipitation

\begin{equation}
R_{\mathrm{eff}} = \eta_r R_f A_{\mathrm{zone}},
\label{eq:effective_rainfall}
\end{equation}

and the net irrigation requirement is

\begin{equation}
D_{\mathrm{net}} = \frac{\max\left(0,D_{\mathrm{req}}-R_{\mathrm{eff}}\right)}{\eta_{\mathrm{irr}}}.
\label{eq:net_irrigation}
\end{equation}

Forecasted rain thus reduces, rather than binarily blocks, irrigation, suppressing it only when net demand reaches zero.

\subsection{Zone-Based Irrigation Decision}

Each zone's pump activates according to

\begin{equation}
I_i =
\begin{cases}
1, & M_i < L \ \text{and}\ D_{\mathrm{net}} > 0,\\
0, & \text{otherwise},
\end{cases}
\label{eq:zone_irrigation}
\end{equation}

The median value is not a direct trigger; a zone can activate even if the overall median is above threshold.

\subsection{Pump Control and Timed Activation}

The BTT Pi V1.2 sends commands over serial to the STM32F103C8T6, which drives relays connected to GPIO pins PB0, PB1, PB10, and PB11:

\begin{equation}
\begin{aligned}
\text{BTT Pi V1.2} &\rightarrow \text{Serial command} \rightarrow \text{STM32F103C8T6} \\
&\rightarrow \text{GPIO} \rightarrow \text{Relay} \rightarrow \text{Pump}.
\end{aligned}
\label{eq:pump_control_path}
\end{equation}

Given net volume $D_{\mathrm{net}}$ and pump flow rate $F_{\mathrm{pump}}$, runtime is

\begin{equation}
t_{\mathrm{pump}} = \max\left(1, \min\left(1800, \frac{D_{\mathrm{net}}}{F_{\mathrm{pump}}} \times 60\right)\right),
\label{eq:pump_duration}
\end{equation}

bounded between 1 s and 30 min, without real-time feedback during watering.

\subsection{Data Acquisition and Monitoring}

Aggregated values at the cycle level and pump summaries (not raw measurements) are stored in a local database, including sensor averages, median values, weather data, and irrigation logs. A PySide6-based graphical user interface displays the most recent 500 data points; statistical analyses of the entire dataset of 38,112 data points (e.g., Pearson correlations) were performed using external tools \cite{sujatha2025gui}.


\section{Implementation}
\label{sec:implementation}

The system was implemented as an embedded edge platform that integrates moisture measurement, sensor processing, weather data retrieval, decision-making, control, and logging, and was tested over a 30-day radish growing period.

\subsection{Hardware Implementation}

The hardware consists of an STM32F103C8T6 (“Blue Pill”), a BTT Pi V1.2, four capacitive soil moisture sensors, a four-channel 5-V relay module, and four DC pumps. The STM32 acquires and forwards the four measured values; the BTT Pi V1.2 calculates the median and performs higher-level processing steps (weather API, decision-making, database, GUI). The most important components are listed in Table~\ref{tab:hardware_components}.

\begin{table}
\caption{Key Hardware Components of the Implemented System}
\label{tab:hardware_components}
\centering
\footnotesize
\setlength{\tabcolsep}{3pt}
\renewcommand{\arraystretch}{1.5}
\begin{tabularx}{\columnwidth}{@{}l l X@{}}
\toprule
\textbf{Component} & \textbf{Model / Type} & \textbf{Function} \\
\midrule
Edge computing unit & BTT Pi V1.2 & High-level processing, control, GUI, weather API, and database management \\
Sensor processing unit & STM32F103C8T6 & Sensor data acquisition \\
Soil moisture sensor & Soil Moisture Sensor V2.0 & Soil moisture measurement \\
Display unit & OLED SSD1306 (I2C) & System status visualization \\
Relay module & 4-channel relay module & Pump circuit and control \\
Irrigation actuators & 4 DC water pumps & Water supply to the growing area \\
\bottomrule
\end{tabularx}
\end{table}

\subsection{Software Implementation}

The Python-based software on the BTT Pi V1.2 coordinates sensor data, weather information, decision-making, actuator control, data storage, and the PySide6 (Qt 6) user interface. The median humidity value is used for monitoring; individual measurements determine decisions at the zone level. Serial commands trigger GPIO signals generated by the STM32 to activate the relays.

\subsection{Terminal Application and Simulation Modes}

A terminal application supports both real and simulated operation (sensors, pumps, weather, number of cycles, interval, mode, database), thereby enabling testing without physical hardware. The STM32 transmits the raw measurement values ($S_1$–$S_4$) via the serial bus at 115,200 baud as CSV files; the BTT Pi calculates the median, thereby separating low-level data acquisition from higher-level processing.

\subsection{Protocol for Serial Communication}

The serial protocol includes handshaking, sequence numbering, acknowledgments, and safety mechanisms, summarized in Table~\ref{tab:serial_protocol}.

\begin{table}
\caption{Serial Communication Protocol and Safety Parameters}
\label{tab:serial_protocol}
\centering
\footnotesize
\setlength{\tabcolsep}{3pt}
\renewcommand{\arraystretch}{1.5}
\begin{tabularx}{\columnwidth}{@{}
>{\raggedright\arraybackslash}p{1.8cm}
>{\raggedright\arraybackslash}p{2.3cm}
>{\raggedright\arraybackslash}X
@{}}
\toprule
\textbf{Parameter} & \textbf{Value} & \textbf{Description} \\
\midrule
Baud rate & 115200 & Serial communication speed \\
Sensor message format & CSV: s1,s2,s3,s4 & Four individual soil moisture measurements \\
Pump command format & PUMP: p1,p2,p3,p4 & Command used to control the four irrigation pumps \\
Pump ACK format & PUMP OK: state1,...,state4 & Acknowledgment message reporting the pump states \\
Heartbeat interval & 5s & Periodic communication message used for connection monitoring \\
Watchdog timeout & 15s & Timeout after which the firmware performs a safety shutdown \\
Handshake & Required at startup & Initial synchronization between the BTT Pi V1.2 and STM32F103C8T6 \\
Safety stop & Automatic on timeout & Watchdog-triggered shutdown of the irrigation pumps \\
\bottomrule
\end{tabularx}
\end{table}

\subsection{Integrating a Weather API}

The controller retrieves a two-day forecast for the configured location. Precipitation and $ET_0$ are included in the irrigation calculations along with the current soil moisture and the crop’s water requirements; temperature is recorded but not used directly. The Python software accesses the API directly, so no separate cloud control server is required.

\subsection{Database and Data Logging}

A local SQLite database logs sensor averages, medians, weather forecasts, irrigation decisions, and irrigation events over a 30-day period and also enables the display of historical data in the graphical user interface.

\subsection{Experimental Application}

The system was operated continuously for 30 days in a large growing container with radish plants, four sensors, and four pumps. During this time, soil moisture readings, weather data, and irrigation events were recorded for later analysis.

\section{Results}
\label{sec:results}

This section describes the experimental setup, methodology, and data analysis based on real-world field measurements, with a focus on soil moisture trends, irrigation events, prevailing weather conditions, and water application rates.

\subsection{Experimental Setup}

The system was evaluated in a one-month cultivation trial with radishes (\textit{Raphanus sativus}) in a large container, with four sensors distributed across the plot and all four pumps in operation. Irrigation decisions were based on measurements from individual zones, plant information, and weather forecasts, while the median supported monitoring. The analysis period spanned from the 13th of one month to the 13th of the following month, during which all subsystems were in continuous operation.

\subsection{Experimental Procedure}

In each cycle, the sensor readings were recorded over a period of 60 s, and time-averaged values and the median were calculated (Section~\ref{sec:methodology}), followed by a 300-s waiting period before the process was repeated. The controller combined the median moisture content, the weather forecast, and the plant parameters; if the moisture content indicated a need for irrigation, effective precipitation reduced this need, and irrigation was activated only if the net demand remained positive, with commands relayed to the pumps via the STM32. All measured values, decisions, and events were continuously logged.

\subsection {Measurements and Analysis}

The most important recorded variables include the time-averaged values per sensor, the median moisture content, weather forecasts, irrigation events, water application rates, and timestamps. The moisture percentages are normalized sensor values derived from a linear mapping (dry = 3000, wet = 1200) and limited to 0–100\%; they represent a normalized output value and not a calibrated volumetric content, as no independent calibration was performed. Soil-moisture-based irrigation planning is well established and can reduce water consumption while maintaining yield \cite{datta2023soil}; accordingly, the median of four sensors—rather than the value from a single sensor—is used here as the primary indicator.

\subsection{Results and Analysis}

\begin{figure}[!t]
\centering
\includegraphics[width=0.9\columnwidth]{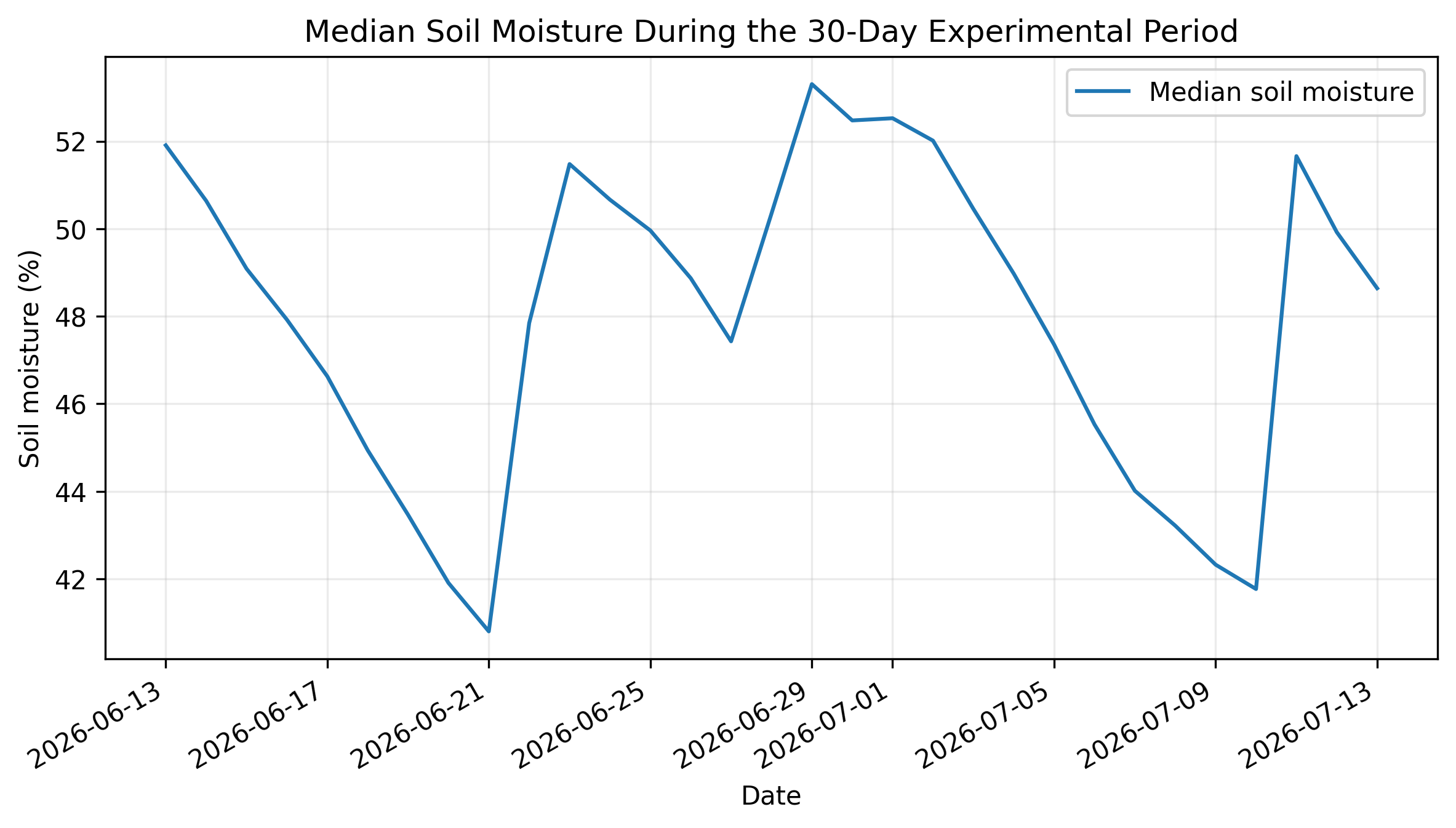}
\caption{Temporal variation in the median soil moisture value during the selected one-month experimental period.}
\label{fig:soil_moisture_results}
\end{figure}

Figure~\ref{fig:soil_moisture_results} shows the median moisture time series over the experimental period, useful for interpreting irrigation events relative to preceding and subsequent conditions.

\begin{figure}
\centering
\includegraphics[width=0.9\columnwidth]{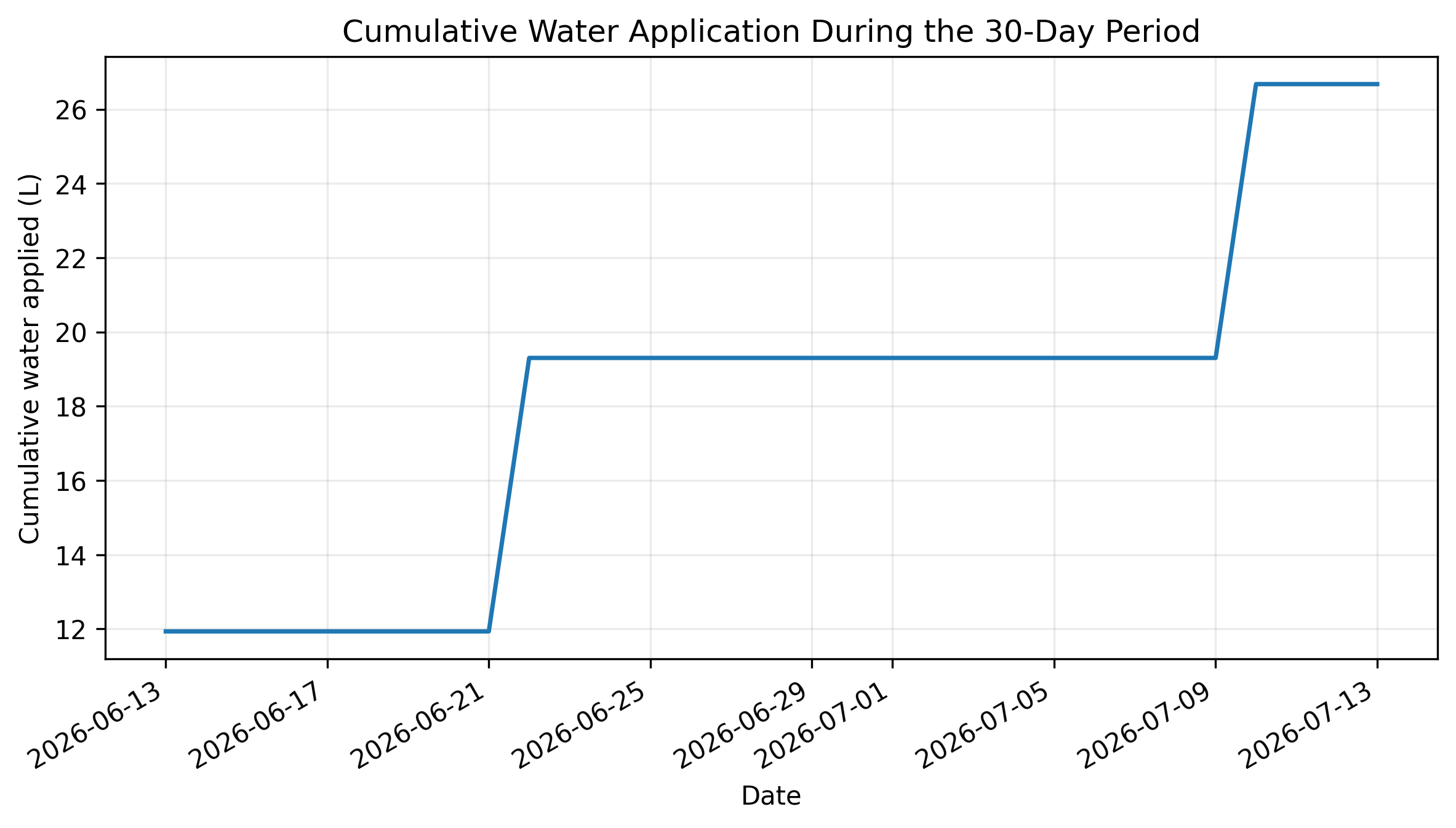}
\caption{Water application recorded during the selected one-month trial period.}
\label{fig:water_application}
\end{figure}

Figure~\ref{fig:water_application} shows the recorded water application, complementing the moisture analysis with a quantitative measure of planned irrigation volumes.

\section{Analysis}
\label{sec:analysis}

This section examines the statistical properties of soil moisture, its relationship to environmental variables, its behavior during irrigation events, the response of soil moisture to irrigation, and the consistency of the sensor measurements.

\subsection{Statistical Characterization of Soil Moisture}

The dataset contains 38,112 aggregated cycle records from June 13 to July 13, 2026 (the sensors recorded data every 2 seconds, averaged over 60-second windows), which were generated using an older control algorithm that differs from the current version. The median soil moisture value was 47.93\%, with a standard deviation of 3.84 percentage points; the values ranged from 32.2\% to 54.0\%. With configured thresholds of 40\% (lower) and 65\% (upper), only 1 out of 38,112 data points fell below the lower threshold (~0.0026\%), and none reached the upper threshold, indicating stable conditions consistent with threshold-based control. The lowest value (32.2\%) triggered irrigation; the highest value (54.0\%) remained below the upper limit. Gradual decreases in moisture between events, followed by increases due to irrigation or rain, suggest that the sensors captured meaningful dynamics of the water balance.

\subsection{Relationship Between Environmental Conditions and Soil Moisture}

The database links soil moisture to $ET_0$, precipitation forecasts (total amounts for the current day and for the next two days), and temperature extremes. The average $ET_0$ value was 4.66 mm/day. The daily Pearson correlations yielded $r=-0.232$ ($p=0.210$) between soil moisture and $ET_0$, as well as $r=0.059$ ($p=0.753$) with the two-day precipitation total—none of the values were statistically significant at $p<0.05$. This does not mean that the weather is irrelevant; soil moisture reflects the combined effects of precipitation, irrigation, evapotranspiration, soil properties, and plant uptake, so correlations with a single variable may be weak even if that variable is physically significant. The stronger (albeit not significant) negative correlation with $ET_0$ is physically consistent with the loss of soil moisture through evaporation; longer data sets would be required for robust significance.

\subsection{Analysis of Irrigation Cycles}

Three irrigation events were recorded, with planned storage volumes of 11.93~L, 7.37~L, and 7.38~L (for a total of 26.68~L)—these are values calculated by the controller and not measured flow rates (the database field for flow rate is set to zero for all older data records). The first irrigation cycle (June 13) took place when soil moisture was at 32.2\% and there was no precipitation forecast. The second cycle (June 22) occurred when soil moisture was 40.0\%, even though there was a precipitation forecast of approximately 0.8 mm (for that day) and 2.1 mm (total over two days), which shows that while weather data is taken into account, it does not override irrigation triggered by threshold values. The third incident (July 10) occurred when soil moisture was 40.0\% with no rain forecast. Conversely, on June 28, a forecast of approximately 22.2 mm was not preceded by any irrigation, and the moisture content rose from 47.4\% to 53.3\% over the following two days, illustrating the substrate’s response to the expected precipitation.

\subsection{Soil Moisture Response to Irrigation}

During the irrigation events on June 22 and July 10, the median soil moisture content rose from approximately 40.1–40.2\% before irrigation to approximately 52.6\% within 30 minutes after irrigation—an increase of about 12.5 percentage points in both cases. This consistency suggests a reproducible short-term relationship between irrigation and measurable changes in moisture, although the lack of pump telemetry means that the physical triggering cannot be independently verified based on the recorded data. The response should be interpreted more as a system-level indicator rather than as a measure of irrigation efficiency or plant water uptake.

\subsection{Water Application and Irrigation Requirements}

The calculated water requirements of the plants for the three events (1.08~L, 0.90~L, 0.96~L) were significantly lower than the planned amounts stored in the old algorithm (11.93~L, 7.37~L, 7.38~L) —which corresponds to ratios of approximately 11.0, 8.2, and 7.7. Applying the current algorithm to the same input values yields values of approximately 4.10~L, 1.27~L, and 1.52~L. Table~\ref{tab:experimental_parameters} lists the parameters used by the current algorithm.

\begin{table}
\caption{Experimental Parameters for the Current Algorithm}
\label{tab:experimental_parameters}
\centering
\footnotesize
\setlength{\tabcolsep}{3pt}
\renewcommand{\arraystretch}{1.5}
\begin{tabularx}{\columnwidth}{@{}l l X@{}}
\toprule
\textbf{Parameter} & \textbf{Value} & \textbf{Description} \\
\midrule
Plant coefficient ($K_c$) & 0.9 & Plant coefficient for radish cultivation \\
Zone area ($A_{\mathrm{zone}}$) & 0.1963 m² & Area of each irrigation zone \\
Effective soil depth ($d$) & 0.3 m & Depth of the effective root zone \\
Lower soil moisture threshold ($L$) & 40\% & Lower threshold for triggering irrigation \\
Upper soil moisture threshold ($U$) & 65\% & Upper threshold for irrigation control \\
Irrigation efficiency ($\eta_{\mathrm{irr}}$) & 0.85 & Efficiency factor for water application \\
Rainfall utilization ($\eta_r$) & 0.80 & Utilization factor for effective precipitation \\
Soil water capacity ($C_w$) & 0.20 & Soil water capacity coefficient \\
Standard pump flow rate & 1.6 L/min & Assumed flow rate for duration calculation \\
\bottomrule
\end{tabularx}
\end{table}

The gap between estimated crop needs and stored planned volumes reflects that physical delivery was never independently measured (delivered-volume field is zero), and that the FAO-56 estimate and the legacy control volume are not expected to coincide exactly. Comparing applied water (26.68~L) to a literature-based radish requirement of about 23.85~L (at $ET_c\approx4.05$~mm/day over 0.1963 m² and 30 days) and to a typical timer-based over-application of 65\% (~39.35~L), the system's application represents roughly 32\% savings relative to conventional timer-based irrigation—though this comparison relies on literature values rather than a controlled in-experiment baseline, since actual delivery was estimated (flow rate $\times$ runtime), not flow-metered.

\subsection{Sensor Agreement and Spatial Variability}

The average moisture content per sensor ranged from 46.28\% (S1) to 48.98\% (S4), corresponding to a standard deviation of approximately 2.70 percentage points, which is consistent with the expected spatial variability of soil properties, water distribution, and sensor placement. The median of the four values, which is used as a representative indicator, provides a robust central estimate that is less sensitive to individual outliers than an arithmetic mean.

\subsection{Statistical Interpretation, Scope, and Future Work}

Since the experiment was conducted in a single growing environment without a parallel control group and only three irrigation events were recorded, the dataset does not allow for statistically sound conclusions regarding water savings or generalizations to other cycles; however, it offers high temporal resolution for characterizing moisture, weather, and dynamics at the event level \cite{ivanova2025soil}. Future work should include a controlled comparison test with timed irrigation, replicates across multiple growth cycles and seasons, and calibration of the assumed pump flow rate of 1.6 L/min using a flow meter to validate the estimated flow rates.

\section{Related Work}
\label{sec:related_work}

In precision agriculture, sensors, communication technology, and computing power are increasingly being integrated to control irrigation, as conventional methods with fixed schedules fail to account for spatial and temporal variations \cite{wolfert2017bigdata,ayaz2019iot}. IoT-enabled smart agriculture features advanced sensor technology, monitoring, and automation \cite{wolfert2017bigdata,ayaz2019iot}, but the heavy reliance on cloud infrastructure leads to latency and a dependence on connectivity \cite{bhuyan2023edge,suganya2023iot}, which motivates the use of edge computing approaches \cite{shi2016edge}.

Wireless soil moisture sensor networks have reduced the effort required for manual monitoring and improved the timing of irrigation \cite{hassan2013webbased,ojha2015wireless,vellidis2008real,eltelbany2019wireless}, but threshold-based methods alone can still result in irrigation occurring shortly before rainfall. The proposed system improves upon this by aggregating data from four spatially distributed sensors using the median and incorporating short-term forecasts to avoid unnecessary irrigation before the predicted rain.

Planning based on weather data and evapotranspiration—in particular, the FAO-56 Penman-Monteith method \cite{allen1998crop} and alternatives such as Hargreaves-Samani and Priestley-Taylor \cite{hargreaves1985reference,priestley1972assessment} —can accurately estimate the water requirements of crops, but in embedded systems, they lead to increased complexity in calculation and calibration \cite{mane2024calibration,jones2004irrigation}. Combining weather data with soil sensors has been shown to improve planning compared to standalone approaches \cite{jamal2023realtime,navarrohellin2015wireless}, and machine learning-based planning is an active area of research \cite{ahmad2025smart}. The proposed system implements a practical, edge-based version of this approach, in which effective precipitation is calculated using short-term forecasts to reduce irrigation requirements.

Reliable embedded hardware that integrates sensors, computing power, communication, and actuators remains indispensable for practical applications \cite{ramesh2021precision}, and edge computing \cite{shi2016edge} enables autonomous operation with low latency. The proposed two-stage design—an STM32F103C8T6 for low-level data acquisition and a BTT Pi V1.2 for data fusion, weather communication, decision-making, database management, and GUI—reduces dependence on the cloud.

Overall, previous work has often focused on individual components (data collection, forecasting, or cloud data management) rather than on comprehensive, field-tested platforms. This paper fills that gap with an integrated, weather-based embedded irrigation system that was evaluated using field data from radish cultivation and combines current soil conditions with forecast information to avoid unnecessary irrigation.

\section{Discussion}
\label{sec:discussion}

The results demonstrate the technical feasibility of integrating multisensor moisture measurement, edge computing, weather data, and automated irrigation into an embedded platform using a sensor-driven paradigm in which weather data supplements the context rather than replacing direct measurements. Using the median of four spatially distributed sensors that showed similar trends but different absolute values yields a fused result that is more robust against outliers than a simple average, while the division of labor between the STM32 and the BTT Pi decouples low-level data acquisition from high-level control.

The irrigation events coincided with low moisture levels, while a major forecasted precipitation event (June 28) was followed by an increase in moisture without triggering irrigation—which is consistent with the fact that forecasts were used as a supplementary factor rather than as the sole determining factor. In the two events that could be analyzed (June 22, July 10), the moisture content rose by about 12.5 points within 30 minutes of irrigation—a repeatable short-term signal, although the lack of telemetry data from the pumps limits an independent verification of the trigger. An important insight from the implementation is the large discrepancy between the calculated water requirements of the crops and the target values stored in the old algorithm, which illustrates that irrigation duration, pump flow rate, and application efficiency must be directly calibrated and cannot be based on assumptions. Similarly, the non-significant correlations between soil moisture and individual weather variables reflect the fact that soil moisture is a state variable influenced by many combined factors—and do not prove that the weather is unimportant.

\begin{figure*}
\centering
\includegraphics[width=0.6\paperwidth]{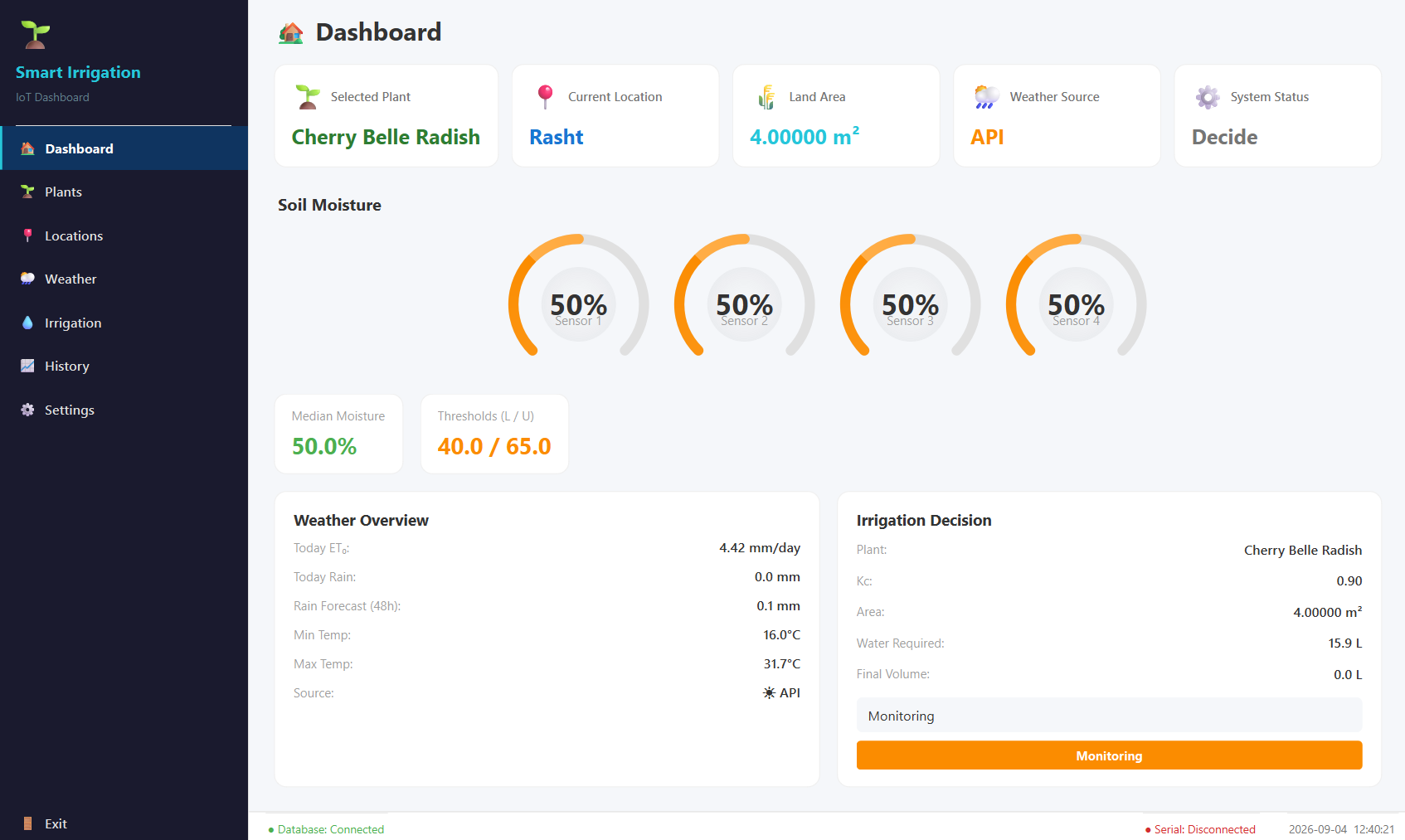}
\caption{Graphical user interface of the implemented irrigation system, displaying real-time sensor readings, weather information, and system status.}
\label{fig:gui}
\end{figure*}

Using local (edge) computing on the BTT Pi V1.2 instead of a remote cloud service reduces dependence on network availability—a valuable advantage for agricultural applications. The associated database and graphical user interface (Figure~\ref{fig:gui}) extend the platform beyond simple pump automation and enable historical analyses as well as operator review without the need to query the database directly.

Despite the robust implementation, there are some limitations to consider. The experimental design, which was limited to a single environment and did not include a control group, does not allow for reliable conclusions regarding the percentage of water savings, yields, or water productivity. Only three irrigation events were recorded, which limits statistical generalizability despite the large number of moisture measurement cycles. In the future, we will expand this work through comprehensive data collection over the course of a year to achieve significantly more precise results. The weather-dependent component also depends on the quality of external forecasts; incorporating forecast reliability rather than a deterministic interpretation of the forecasts could improve robustness.

Notwithstanding these limitations, the results provide a practical foundation for further development. The modular combination of decentralized sensors, edge fusion, weather-aware decision-making, automated control, and continuous logging can be extended to other crops and environments, thereby supporting the overarching goal of scalable platforms for smart agriculture. Future work will expand data collection across multiple growing seasons and investigate machine learning-based predictive planning.

\section{Conclusion}
\label{sec:conclusion}

This article presents the design, implementation, and experimental evaluation of an embedded, weather-based irrigation system that integrates soil moisture monitoring using multiple sensors, edge computing, weather forecast data, automated control, and local data management. Four capacitive sensors connected to an STM32F103C8T6 collect data, while a BTT Pi V1.2 edge platform handles decision-making, weather data retrieval, and database management, thereby separating low-level measurement from higher-level control. During a one-month radish growing period (June 13 to July 13, 2026), 38,112 aggregated measurement cycle data records were collected. The median soil moisture level averaged 47.93\% (standard deviation 3.84 points, range 32.2–54.0\%). Three irrigation events resulted in a planned total volume of 26.68 L (calculated by the controller, not physically measured); the two events with a valid pre-irrigation reference value showed an increase in moisture of approximately 12.5 percentage points shortly after irrigation. These results demonstrate the technical feasibility of integrating environmental measurements, weather data, local decision-making, physical control, and continuous data collection into a single embedded irrigation platform, thereby laying the foundation for future analyses and next-generation precision agriculture systems.

\section{Acknowledgment}

The authors note that they used Grammarly and AI-powered language tools to improve the manuscript’s grammar, linguistic clarity, and overall readability. The authors independently reviewed, revised, and verified the final manuscript and assume full responsibility for its content, accuracy, and integrity. The authors intend to make the irrigation system they developed available as an open-source project following publication.

\bibliographystyle{IEEEtran}
\bibliography{references}

@article{shakeri2018uav,
  author = {Shakeri, Reza and Al-Garadi, Mohammed Ali and Badawy, Ahmed and Mohamed, Amr and Khattab, Tamer and Al-Ali, Abdulla Khalid and Harras, Khaled A. and Guizani, Mohsen},
  title = {Design Challenges of Multi-UAV Systems in Cyber-Physical Applications: A Comprehensive Survey and Future Directions},
  journal = {IEEE Communications Surveys and Tutorials},
  volume = {21},
  number = {4},
  pages = {3340-3385},
  year = {2018},
  publisher = {Institute of Electrical and Electronics Engineers (IEEE)},
  doi = {10.1109/comst.2019.2924143},
}

@article{wolfert2017bigdata,
  author = {S. Wolfert and L. Ge and C. Verdouw and M.-J. Bogaardt},
  title = {Big Data in Smart Farming -- A Review},
  journal = {Agricultural Systems},
  volume = {153},
  pages = {69--80},
  year = {2017},
  publisher = {Elsevier BV},
  doi = {10.1016/J.AGSY.2017.01.023},
}

@article{ayaz2019iot,
  author = {M. Ayaz and M. Ammad-Uddin and Z. Sharif and A. Mansour and E.-H. M. Aggoune},
  title = {Internet-of-Things (IoT)-Based Smart Agriculture: Toward Making the Fields Talk},
  journal = {IEEE Access},
  volume = {7},
  pages = {129551--129583},
  year = {2019},
  publisher = {Institute of Electrical and Electronics Engineers (IEEE)},
  doi = {10.1109/ACCESS.2019.2932609},
}

@article{shi2016edge,
  author = {W. Shi and J. Cao and Q. Zhang and Y. Li and L. Xu},
  title = {Edge Computing: Vision and Challenges},
  journal = {IEEE Internet of Things Journal},
  volume = {3},
  number = {5},
  pages = {637--646},
  year = {2016},
  publisher = {Institute of Electrical and Electronics Engineers (IEEE)},
  doi = {10.1109/JIOT.2016.2579198},
}

@article{vellidis2008real,
  author = {G. Vellidis and M. Tucker and C. Perry and C. Kvien and C. Bednarz},
  title = {A Real-Time Wireless Smart Sensor Array for Scheduling Irrigation},
  journal = {Computers and Electronics in Agriculture},
  volume = {61},
  number = {1},
  pages = {44--50},
  year = {2008},
  publisher = {Elsevier BV},
  doi = {10.1016/J.COMPAG.2007.05.009},
}

@article{eltelbany2019wireless,
  author = {El-Telbany, M. and Maged, Maha A.},
  title = {Wireless sensor networks for extreme environments: remote sensing and space industry},
  journal = {International Journal of Hybrid Intelligence},
  volume = {1},
  number = {1},
  pages = {41},
  year = {2019},
  publisher = {Inderscience Publishers},
  doi = {10.1504/IJHI.2019.10021295},
}

@article{jones2004irrigation,
  author = {H. G. Jones},
  title = {Irrigation Scheduling: Advantages and Pitfalls of Plant-Based Methods},
  journal = {Journal of Experimental Botany},
  volume = {55},
  number = {407},
  pages = {2427--2436},
  year = {2004},
  publisher = {Oxford University Press (OUP)},
  doi = {10.1093/JXB/ERH213},
}

@article{navarrohellin2016decision,
  author = {Navarro-Hell{\'i}n, H. and Rinc{\'o}n, J. M. D. and Domingo-Miguel, R. and Soto-Vall{\'e}s, F. and Torres-S{\'a}nchez, R.},
  title = {A Decision Support System for Managing Irrigation in Agriculture},
  journal = {Computers and Electronics in Agriculture},
  volume = {124},
  pages = {121-131},
  year = {2016},
  publisher = {Elsevier BV},
  doi = {10.1016/J.COMPAG.2016.04.003},
}

@article{sui2017irrigation,
  author = {Sui, Ruixiu},
  title = {Irrigation Scheduling Using Soil Moisture Sensors},
  journal = {Journal of Agricultural Science},
  volume = {10},
  pages = {1},
  year = {2017},
  publisher = {Canadian Center of Science and Education},
  number = {1},
  doi = {10.5539/jas.v10n1p1},
}

@article{ojha2015wireless,
  author = {T. Ojha and S. Misra and N. S. Raghuwanshi},
  title = {Wireless Sensor Networks for Agriculture: The State-of-the-Art in Practice and Future Challenges},
  journal = {Computers and Electronics in Agriculture},
  volume = {118},
  pages = {66--84},
  year = {2015},
  publisher = {Elsevier BV},
  doi = {10.1016/j.compag.2015.08.011},
}

@article{jamal2023realtime,
  author = {Jamal, A. and Cai, X. and Qiao, X. and Garc{\'i}a, L. and Wang, J. and Amori, A. and Yang, H.},
  title = {Real-Time Irrigation Scheduling Based on Weather Forecasts, Field Observations, and Human-Machine Interactions},
  journal = {Water Resources Research},
  volume = {59},
  pages = {25--35},
  year = {2023},
  publisher = {American Geophysical Union (AGU)},
  doi = {10.1029/2023WR035810},
  number = {12},
}

@article{navarrohellin2015wireless,
  author = {J. M. Navarro-Hell{\'i}n and R. Torres-S{\'a}nchez and F. Soto-Valles and C. Albaladejo-P{\'e}rez and J. A. L{\'o}pez-Riquelme and R. Domingo-Miguel},
  title = {A wireless sensors architecture for efficient irrigation management},
  journal = {Agricultural Water Management},
  volume = {151},
  pages = {64-74},
  year = {2015},
  publisher = {Elsevier BV},
  doi = {10.1016/j.agwat.2014.10.022},
}

@article{datta2023soil,
  author = {S. Datta and S. Taghvaeian},
  title = {Soil Water Sensors for Irrigation Scheduling in the United States: A Systematic Review of Literature},
  journal = {Agricultural Water Management},
  volume = {278},
  pages = {108148},
  year = {2023},
  publisher = {Elsevier BV},
  doi = {10.1016/j.agwat.2023.108148},
}

@article{mane2024calibration,
  author = {S. Mane and N. Das and G. Singh and M. Cosh and Y. Dong},
  title = {Advancements in Dielectric Soil Moisture Sensor Calibration: A Comprehensive Review of Methods and Techniques},
  journal = {Computers and Electronics in Agriculture},
  volume = {218},
  pages = {108686},
  year = {2024},
  doi = {10.1016/j.compag.2024.108686},
  publisher = {Elsevier BV},
}

@article{pereira2021cropcoefficients,
  author = {Pereira, L. S. and Paredes, P. and Hunsaker, D. and L{\'o}pez-Urrea, R. and Shad, Z. M.},
  title = {Standard Single and Basal Crop Coefficients for Field Crops: Updates and Advances to the FAO56 Crop Water Requirements Method},
  journal = {Agricultural Water Management},
  volume = {243},
  pages = {106466},
  year = {2021},
  doi = {10.1016/j.agwat.2020.106466},
  publisher = {Elsevier BV},
}

@article{ivanova2025soil,
  author = {I. Ivanova},
  title = {Soil Moisture Forecasting from Sensors-Based Soil Moisture, Weather and Irrigation Observations: A Systematic Review},
  journal = {Smart Agricultural Technology},
  volume = {10},
  pages = {100692},
  year = {2025},
  doi = {10.1016/j.atech.2024.100692},
  publisher = {Elsevier BV},
}

@article{postel1996human,
  author = {Postel, S. and Daily, G. and Ehrlich, P.},
  title = {Human appropriation of renewable fresh water},
  journal = {Science},
  volume = {271},
  number = {5250},
  pages = {785--788},
  year = {1996},
  publisher = {American Association for the Advancement of Science (AAAS)},
  doi = {10.1126/science.271.5250.785},
}

@article{mahajan2025precision,
  author = {Mahajan, Rupali Atul and Dey, Rajesh and Shinde, Gitanjali and Bhuyan, Basanta Kumar and Bhuyan, Kalpana and Desai, Sharada},
  title = {Precision Irrigation: A New Approach to Water Management in Agriculture},
  journal = {International Journal of Environment and Waste Management},
  volume = {1},
  number = {1},
  pages = {420-432},
  year = {2025},
  publisher = {Inderscience Publishers},
  doi = {10.1504/ijewm.2025.10070501},
}

@inproceedings{muzdrikah2018calibration,
  title={Calibration of capacitive soil moisture sensor (SKU: SEN0193)},
  author={Muzdrikah, Fajar Siti and Nuha, M Shohibun and Rizqi, Fathi A and others},
  booktitle={2018 4th International Conference on Science and Technology (ICST)},
  pages={1--6},
  year={2018},
  organization={IEEE}
}

@article{khasawneh2026optimizing,
  author = {Khasawneh, Hussam J. and Hammad, Nour and Al-Sweity, Dana},
  title = {Optimizing Water Use with an Arduino IoT-Based Smart Irrigation System},
  journal = {Advances in Transdisciplinary Engineering},
  volume = {155},
  pages = {64--71},
  year = {2026},
  publisher = {IOS Press},
  doi = {10.3233/atde260065},
}

@article{ahmad2025smart,
  author = {Ahmad, Alaa Yahya Aref and Ma’aitah, Mohammad K. S.},
  title = {Smart Irrigation Prediction Using Machine Learning},
  journal = {2025 International Conference on Business Intelligence for Technology Innovation (ICBITI)},
  volume = {9},
  number = {6},
  pages = {1-8},
  year = {2025},
  publisher = {IEEE},
  doi = {10.1109/icbiti65527.2025.11500621},
}

@article{bhuyan2023edge,
  author = {Bhuyan, Bikram Pratim and Tomar, Ravi and Yadav, Anuj Kumar},
  title = {Edge Computing in Smart Agriculture},
  journal = {Future Connected Technologies},
  volume = {4},
  number = {5},
  pages = {200-217},
  year = {2023},
  publisher = {CRC Press},
  doi = {10.1201/9781003287612-10},
}

@article{suganya2023iot,
  author = {Suganya, S. and Kowsalya, K. and Abdul Hayum, A. and Latha, P.},
  title = {IOT-BASED SMART AGRICULTURE WITH CLOUD COMPUTING},
  journal = {IIP Series},
  volume = {10},
  number = {3},
  pages = {103-116},
  year = {2023},
  publisher = {Iterative International Publishers (IIP)},
  doi = {10.58532/v3bdio7p2ch7},
}

@article{hassan2013webbased,
  author = {Hassan, Ayman M.},
  title = {Web-based irrigation management for open canals using Wireless Sensor Networks},
  journal = {2013 IEEE Conference on Wireless Sensor (ICWISE)},
  volume = {19},
  number = {3},
  pages = {102-107},
  year = {2013},
  publisher = {IEEE},
  doi = {10.1109/icwise.2013.6728789},
}

@article{ramesh2021precision,
  author = {Ramesh, S. and Kumar, M. Vinoth},
  title = {A Review of Precision Agriculture Based on Embedded System Applications},
  journal = {Artificial Intelligence and Cybersecurity},
  volume = {12},
  number = {3},
  pages = {87-106},
  year = {2021},
  doi = {10.1201/9781003097518-6},
  publisher = {CRC Press},
}

@article{sujatha2025gui,
  author = {S., Dr. Sujatha},
  title = {Design and implementation of GUI for IoT-based agricultural monitoring systems},
  journal = {International Journal of Innovative Research in Information Security},
  volume = {11},
  pages = {818},
  year = {2025},
  publisher = {AM Publications},
  number = {11},
  doi = {10.26562/ijiris.2025.v1111.19},
}

@article{hargreaves1985reference,
  author = {G. H. Hargreaves and Z. A. Samani},
  title = {Reference crop evapotranspiration from temperature},
  journal = {Applied Engineering in Agriculture},
  volume = {1},
  number = {2},
  pages = {96--99},
  year = {1985},
  publisher = {American Society of Agricultural and Biological Engineers (ASABE)},
  doi = {10.13031/2013.26773},
}

@article{priestley1972assessment,
  author = {C. H. B. Priestley and R. J. Taylor},
  title = {On the assessment of surface heat flux and evaporation using large-scale parameters},
  journal = {Monthly Weather Review},
  volume = {100},
  number = {2},
  pages = {81--92},
  year = {1972},
  publisher = {American Meteorological Society},
  doi = {10.1175/1520-0493(1972)100<0081:OTAOSH>2.3.CO;2},
}

@article{anbarasu2026multiparameter,
  author = {Anbarasu, R. and Parthipan, V.},
  title = {Multi-Parameter Soil Moisture Estimation Using Low-Cost Sensor Fusion for Precision Irrigation},
  journal = {2026 International Conference on Computational Robotics, Testing and Engineering Evaluation (ICCRTEE)},
  volume = {13},
  number = {4},
  pages = {1-6},
  year = {2026},
  publisher = {IEEE},
  doi = {10.1109/ICCRTEE68719.2026.11566518},
}

@article{wang2016datafusion,
  author = {Wang, Man-tao and Wei, Jiang-shu and Pan, Yong-hao and Wei, Zhe},
  title = {Study on Data Fusion Techniques in Wireless Sensor Networks},
  journal = {Proceedings of the 6th International Asia Conference on Industrial Engineering and Management Innovation},
  volume = {150},
  pages = {67-74},
  year = {2016},
  publisher = {Atlantis Press},
  doi = {10.2991/978-94-6239-145-1_7},
}

@article{macadam2020irrigation,
  author = {MacAdam, J. and Barta, A. L.},
  title = {Irrigation and water management.},
  publisher = {CABI},
  year = {2020},
  journal = {Turfgrass physiology and ecology: advanced management principles},
  pages = {107-128},
  doi = {10.1002/9781119436669.CH27},
}

@article{rai2017efficiency,
  author = {Rai, R. K. and Singh, V. and Upadhyay, A.},
  title = {Irrigation efficiency},
  journal = {Irrigation Science},
  volume = {22},
  number = {2},
  pages = {55--61},
  year = {2017},
  publisher = {Springer},
  doi = {10.1016/B978-0-12-811748-4.00018-2},
}

@book{allen1998crop,
author = {R. G. Allen and L. S. Pereira and D. Raes and M. Smith},
title = {Crop Evapotranspiration: Guidelines for Computing Crop Water Requirements},
publisher = {FAO Irrigation and Drainage Paper 56},
year = {1998},
address = {Rome, Italy},
isbn = {92-5-104219-5},
url = {https://www.fao.org/4/x0490e/x0490e00.htm},
}

@article{unwater2020climate,
title = {Water and climate change},
author = {Water, UN},
journal = {The United Nations World Water Development Report},
year = {2020},
publisher = {UNESCO, Paris, France},
url = {https://www.unwater.org/publications/un-world-water-development-report-2020},
}

\end{document}